\documentclass[aps,nofootinbib,preprintnumbers,nobalancelastpage,superscriptaddress]{revtex4}

\usepackage[english]{babel}
\usepackage{amsmath,amssymb}
\usepackage[pdftex]{graphicx}
\usepackage[sort&compress]{natbib}
\usepackage{amsfonts}
\usepackage{bbm}
\usepackage{color}

\newcommand{\bc}{\begin{center}}
\newcommand{\ec}{\end{center}}
\newcommand{\bt}{\begin{tabular}}
\newcommand{\et}{\end{tabular}}
\newcommand{\be}{\begin{equation}}
\newcommand{\ee}{\end{equation}}
\newcommand{\bea}{\begin{eqnarray}}
\newcommand{\eea}{\end{eqnarray}}
\newcommand{\bfig}{\begin{figure}}
\newcommand{\efig}{\end{figure}}

\def\gsim{\mathrel{\lower3pt\hbox{$\sim$}}\hskip-11.5pt\raise3pt\hbox{$>$}\;}

\def\lsim{\mathrel{\lower3pt\hbox{$\sim$}}\hskip-11.5pt\raise3pt\hbox{$<$}\;}

\newcommand{\AddrULB}{%
 Service de Physique Th\'eorique, Universit\'e Libre de Bruxelles,\\
CP225, Bld du Triomphe, 1050 Brussels, Belgium}

\newcommand{\AddrDESY}{%
Deutsches Elektronen Synchrotron,
Notkestrasse 85,
D-22603 Hamburg, Germany}

\newcommand{\AddrDurham}{%
Institute for Particle Physics Phenomenology,
Durham University,
DH1 3LE, UK
}

\begin{document}

\preprint{ULB-TH/10-05}
\preprint{DESY 10-036}

\title{
A light scalar WIMP through the Higgs portal  and CoGeNT}

\vspace{3mm}
\author{Sarah Andreas}\affiliation{\AddrDESY}
\author{Chiara Arina} \affiliation{\AddrULB}
\author{Thomas Hambye}\affiliation{\AddrULB}
\author{ Fu-Sin Ling}\affiliation{\AddrULB}\affiliation{\AddrDurham}
\author{Michel H.G.~Tytgat}\affiliation{\AddrULB}

\begin{abstract}
If dark matter (DM) simply consists in a scalar particle interacting dominantly with the Higgs boson, the ratio of its annihilation cross section ---which is relevant both for the relic abundance and  indirect detection--- and its spin-independent scattering cross section on nuclei depends only on the DM mass. It is an intriguing result that, fixing the mass and direct detection rate to fit the annual modulation observed by the DAMA experiment,
one obtains a relic density in perfect agreement with its observed value. In this short letter we update this result and confront the model to the recent CoGeNT data, tentatively interpreting the excess of events in the recoil energy spectrum as being due to DM.
CoGeNT, as DAMA, points toward a light DM candidate, with somewhat different (but not necessarily incompatible) masses and cross sections. For the CoGeNT region too, we find an intriguing agreement between the scalar DM relic density and direct detection constraints. 
We  give the one $\sigma$ region favoured by the CDMS-II events, and our exclusion limits for the Xenon10 (2009) and Xenon100 data, which, depending on the scintillation efficiency, may exclude CoGeNT and DAMA.
Assuming CoGeNT and/or DAMA to be due to scalar singlet DM leads to definite predictions regarding indirect detection and at colliders. We specifically  emphasize the limit on the model that might be set by the current {\it Fermi}-LAT data on dwarf galaxies, and the implications for the search for the Higgs at the LHC.
\end{abstract}
\maketitle


\section{Introduction}

Recently, there has been  some effervescence regarding what may be first hints of direct detection of dark matter (DM) from the galactic halo. The most recent is related to the CoGeNT experiment, a low background germanium crystal based detector, with a rather modest exposure time and detector mass, but very low threshold energy, which has announced an anomaly in the form of an excess of events at low recoil energies  \cite{Aalseth:2010vx}. Although the collaboration clearly leans toward natural radioactivity as the cause for the observed excess, they do put forward the possibility that the events may be explained by the elastic collisions of DM from the galactic halo, with a mass in the $\sim 7-10$ GeV range, and a rather large spin independent (SI) cross section on nuclei, $\sigma_{SI} \sim 7 \times 10^{-41}$ cm$^{2}$. 

Surprizingly, these values for the mass and scattering cross section are not too different from those required to fit the DAMA/Libra and DAMA/NaI (DAMA in the sequel) events. DAMA has  observed 11 successive cycles 
of annual modulation in the rate of nuclear recoils, with a statistical significance of $8.2 \sigma$ \cite{Bernabei:2008yi}. 
These measurements are consistent with the signal that would arise from elastic scattering of a WIMP from the galactic halo with the nuclei in the detectors, the flux of DM particles being  modulated by the 
periodic motion of the Earth around the Sun~\cite{Drukier:1986tm,Freese:1987wu}.

There has been much work on the DM interpretation of the recent DAMA data (see {\em e.g.} \cite{Bottino:2008mf,Foot:2008nw,Chang:2008xa,Khlopov:2008ki,Bernabei:2008mv,Feng:2008dz,Masso:2009mu,Petriello:2008jj,Savage:2008er,Chang:2008gd,Dudas:2008eq,Cui:2009xq,MarchRussell:2008dy,Fairbairn:2008gz,Bandyopadhyay:2010cc,Ling:2009cn}). 
In  \cite{Andreas:2008xy} (see also \cite{Arina:2009um}), it has been shown that the DAMA results may be explained as being caused by the elastic scattering  of a  singlet scalar DM candidate interacting through the Higgs portal. This particle may be a true singlet scalar \cite{McDonald:1993ex,Burgess:2000yq,Barger:2007im}, or the low energy limit of another model\footnote{A possible implementation is the Inert Doublet Model (IDM), which is another very simple extension of the Standard Model with scalar DM \cite{Deshpande:1977rw,Barbieri:2006dq,Ma:2006km,LopezHonorez:2006gr}.}.

\section{Confrontation to CoGeNT}

In the present short letter, we 
provide an update on the status of the singlet scalar in the light of the recent data from the CoGeNT collaboration, tentatively interpreted as being due to DM. We also include the recent data of the CMDS-II collaboration \cite{Ahmed:2009zw}. With 2 events for an expected background of $0.8 \pm 0.2$ events, the significance of this result is low, but it also hints at a rather light DM candidate.

We adopt the convention of \cite{Andreas:2008xy}, and consider just one real singlet scalar $S$, together with a $Z_2$ symmetry, $S \rightarrow -S$, to ensure its stability, so that the following renormalizable terms may be added to the SM Lagrangian:
\begin{equation}
\label{lag}
{\cal L} \owns \frac{1}{2}\partial^\mu S \partial_\mu S-\frac{1}{2}\mu^2_S \,S^2 -\frac{\lambda_S}{4} S^4 -\lambda_L\, H^\dagger H\, S^2
\end{equation} 
where $H$ is the Standard Model Higgs doublet, and the mass of $S$ is  given by 
\begin{equation}
\label{massS}
m_S^2=\mu^2_S+\lambda_L \mbox{\rm v}^2
\end{equation}
with $\hbox{v}=246$~GeV.   
Both annihilation into SM particles  and scattering with nuclei are through the coupling $\lambda_L$ to the Higgs particle $h$, respectively in the s and  t-channel. Annihilation through the Higgs is S-wave, and  the cross section for scattering on nuclei is purely of the  SI type. For a DM candidate lighter than the Higgs, $m_S \ll m_h$, the ratio of the annihilation and scattering cross section depends only on $m_S$,
\begin{equation}
\sum_{\mbox{\rm \small f}} \frac{\sigma(S S \rightarrow \bar{\mbox{\rm f}} \mbox{\rm f}) v_{rel}}{\sigma(S N \rightarrow  S N)}
=\sum_{\mbox{\rm \small f}}
\frac{n_c m^2_{\mbox{\rm f}}}{f^2 m^2_N \mu^2_r}
\frac{(m_S^2-m_{\mbox{\rm f}}^2)^{3/2}}{m_S}
\label{ratioS}
\end{equation}
where ${n_c}=3(1)$ for quarks (leptons),  and $\mu_r=m_S m_{N}/(m_S+m_{N})$ is the nucleon-DM reduced mass. The factor $f$ parametrizes the Higgs to nucleons coupling, $f m_{N}\equiv \langle {N}| \sum_q m_q \bar{q}q|{N}\rangle=g_{h{NN}} \mbox{\rm v}$, and  we consider $0.2 \leq f\leq 0.4$ (see {\em e.g.} \cite{Gasser:1990ap}).

\begin{figure}
\includegraphics[width=0.5\columnwidth]{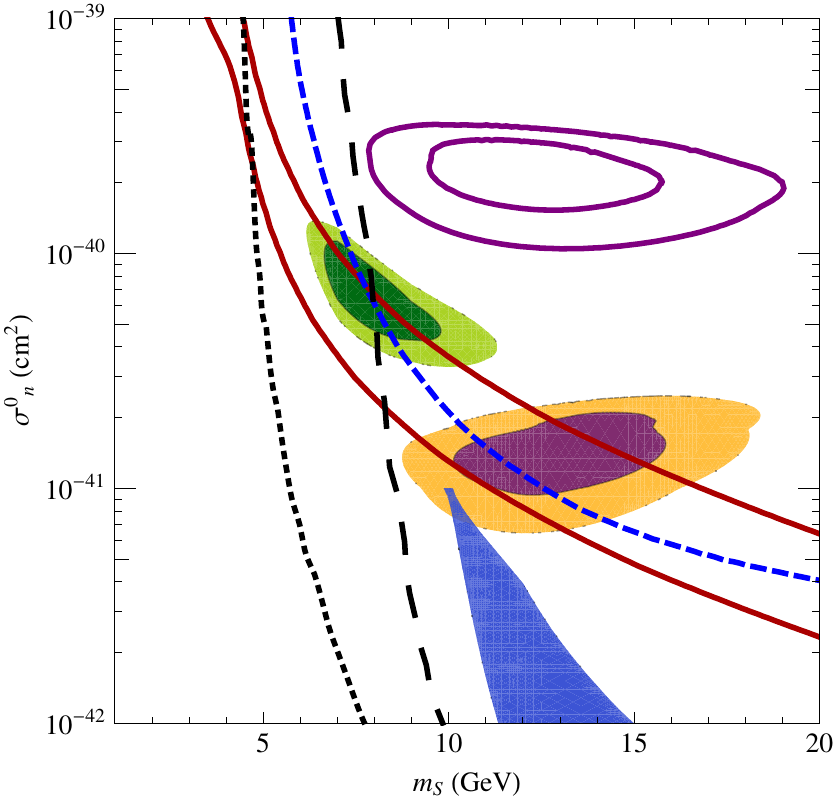}
\caption{SI cross section ($\sigma^0_{n}$) {\em vs} scalar singlet mass ($m_S$), for $\rho_{DM} = 0.3$ GeV/cm$^3$ and a standard Maxwellian velocity distribution (with mean velocity $220$ km/s and escape velocity $v_{esc}=650$ km/s, see our conventions in \cite{Arina:2009um}). The green region corresponds to CoGeNT (minimum $\chi^2$, with contours at 90 and 99.9\% C.L.), for which we have assumed that the excess at low recoil energies is entirely due to DM (assuming a constant background contamination). The DAMA regions (goodness-of-fit, also at 90 and 99.9\% C.L.) are given both with (purple/orange) and without (purple, no fill) channelling. The blue region corresponds to the CDMS-II two events, at $1 \sigma$, which we obtained following the procedure of \cite{Kopp:2009qt}.   The blue (short-dashed) line is the 90\% C.L. exclusion limit from CDMS-Si \cite{Akerib:2005kh}. The  black dotted line is the $90 \%$ C.L. exclusion limit from the Xenon10 2009 data set, using their scintillation efficiency \cite{Angle:2009xb}, as also considered in \cite{Kopp:2009qt}. The long-dashed line is based on the same data but using instead the  smaller scintillation efficiency advocated in \cite{Manzur:2009hp} (central value, at $1\sigma$ the corresponding exclusion can be found in \cite{Fitzpatrick:2010em}).
Finally, the brown lines (continuous) encompass the region predicted by the singlet scalar DM model corresponding to the WMAP range $0.094 \leq \Omega_{DM} h^2 \leq 0.129$, for $0.2 \leq f \leq 0.4$.}
\label{fig:elastic_update}
\end{figure}

Equation (\ref{ratioS}) shows that the mass of the DM candidate is fixed for a given relic abundance and SI scattering cross section \cite{McDonald:1993ex,Burgess:2000yq,Barger:2007im,Andreas:2008xy}. In turn, direct detection experiments may determine both the SI cross section  and the mass of the DM, modulo the astrophysical uncertainties regarding the local density and  velocity distribution of the DM. {\em A priori} there is little chance that these constraints may be met by singlet scalar DM, but as Figure \ref{fig:elastic_update} reveals, the model may be in agreement  with CoGeNT ---which is the main result of this letter--- or, as shown in \cite{Andreas:2008xy,Arina:2009um}, with DAMA. Since there is a gap between the CoGeNT and DAMA (with channelling) regions, that scalar DM agrees with CoGeNT does not trivially derive from the fact that it may agree with DAMA. We emphasize that this result, as for DAMA, is specific to a scalar particle with scalar couplings to SM fermions. For instance, annihilation through the Higgs portal would be P-wave suppressed for a fermionic singlet DM candidate,  and other interactions, as is the case  for a  light neutralino \cite{Bottino:2007qg}, are necessary to agree with the direct detection data (see also {\em e.g.} \cite{Kim:2009ke,Fitzpatrick:2010em,Kuflik:2010ah,Feldman:2010ke}).

The gap between CoGeNT and DAMA  may be reduced, either assuming that channelling is less effective than what is advocated by the DAMA collaboration (which has the effect of raising the DAMA region --- but not reducing the tension with exclusion limits), or  assuming that the CoGeNT excess is partially contaminated by some natural radioactivity (lowering the CoGeNT region) or a mixture of both. One may also adjust the properties of the halo or the DAMA spectral data \cite{Fitzpatrick:2010em}, but we have refrained from doing so.

Also, both regions are excluded by the most stringent limit set by Xenon10 (the dotted line in Figure \ref{fig:elastic_update}). However, if the scintillation efficiency ---the measure of the fraction of energy from the recoiling nuclei that goes into scintillation light---  is actually lower than that used by the collaboration (the long-dashed line in Figure  \ref{fig:elastic_update}), as advocated in \cite{Manzur:2009hp}, there is a region of CoGeNT which may be consistent with all experimental constraints (and the singlet scalar DM candidate).

\begin{figure}[t!]
\centering
\includegraphics[width=0.450\columnwidth]{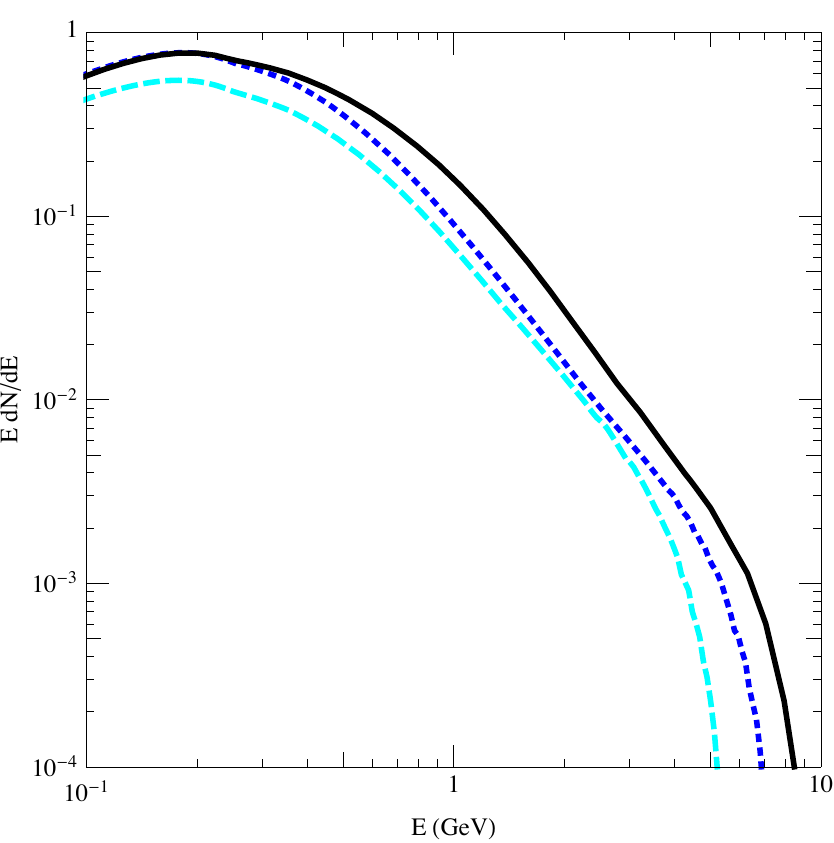}
\caption{$E dN/dE$ spectra produced by Pythia8.1 for a scalar singlet of mass $6$ GeV (blue, long dashed), $8$ GeV (dark blue, short dashed) and $10$ GeV (black, continuous line).}
\label{fig:spectra}
\end{figure}

In the scenario we consider, the coupling $\lambda_L$ to the Higgs must be fairly large to explain both the relic abundance\footnote{In the calculation of the relic abundance, we neglect the possible effect of the QCD phase transition, which is supposed to have taken place around  $\sim 150$ MeV (see, for instance, \cite{McLerran:2008ux}).
Since typically $x_f =m_S/T_{fo} \simeq 20$,  for $m_S \simeq 6$ GeV, for instance, the freeze-out temperature  is   $T_{fo} \simeq 300$ MeV $\gg 150$ MeV and thus the QCD phase transition is irrelevant for the range of mass we consider. Otherwise, the QCD phase transition  might increase the relic abundance by at most a factor of {\cal O}(2) (see for instance \cite{Bottino:2003iu}). Such an increase could be compensated by an increase of $\lambda_L$ by a factor of $\sim 1.4$ and this, in turn, would require a decrease of the parameter $f$ by the same factor.}
 and the direct detection data. For $m_h = 120$ GeV ($m_h = 180$ GeV), one typically requires $\lambda_L \simeq -0.2$ (resp. $\simeq -0.45$) for a DM candidate of mass $m_S \sim 8$ GeV. 
For the same choice of parameters, $\mu_S \sim 100$ GeV, which implies  fine tuning of the parameters at the level of the percent, which  is not unbearable in our opinion. Also there is no  mechanism to naturally stabilize the mass of the scalar at a scale of few GeV. Note however that, if neither CoGeNT nor DAMA could be fitted with $\mu_S^2 \sim 0$ at tree level, in the Inert Doublet Model  at one loop,  the DM mass and coupling ranges required by CoGeNT and/or DAMA may be compatible with dynamical electroweak symmetry breaking induced by the inert doublet \cite{Hambye:2007vf}.

A light scalar dark matter candidate coupled to the Higgs  has potentially many other signatures or implications, a  large flux of gamma rays from dark matter annihilations \cite{Feng:2008dz,Andreas:2008xy}, a large flux of neutrinos from capture by the Sun, which may be constrained by Super-Kamiokande \cite{Savage:2008er,Savage:2009mk,Andreas:2009hj,Feng:2008qn}, or anti-protons and anti-deuterons in cosmic rays \cite{Bottino:2008mf,Nezri:2009jd}. 

\begin{table}[t]
  \begin{center}
    \begin{tabular}{c|c|c|c|c}
& \multicolumn{2}{|c|}{Ursa Minor} & \multicolumn{2}{|c}{Draco }  \\
       \hline 
       $m_S$ and BR & $\Phi_{\rm pred} (\rm cm^{-2} s^{-1})$ &$\Phi_{\rm lim}^{95\% \rm CL} (\rm cm^{-2} s^{-1})$ &$\Phi_{\rm pred} (\rm cm^{-2} s^{-1})$ & $\Phi_{\rm lim}^{95\% \rm CL} (\rm cm^{-2} s^{-1})$ \\
       
      \hline\hline
      10 GeV &   & & & \\
       BR($SS \rightarrow \tau^+\tau^-$) $\simeq 10\%$ &$8.5 \times 10^{-10}$ &$7.8 \times 10^{-10}$ & $1.6 \times 10^{-9}$ & $1.6 \times 10^{-9}$\\
      BR($SS \rightarrow b\bar{b}+ c\bar{c}$) $\simeq 90\%$ & & & &\\ 
      \hline
       6 GeV & & & &\\ 
      BR($SS \rightarrow \tau^+\tau^-$) $\simeq 20\%$ &$1.5 \times 10^{-9}$ &$1.0\times 10^{-9}$ & $2.8 \times 10^{-9}$&$1.7\times 10^{-9}$ \\	
      BR($SS \rightarrow b\bar{b}+ c\bar{c}$) $\simeq 80\%$ & & & & \\ 
      \hline\hline
    \end{tabular}    
       \label{tab:dwarfs}
  \end{center}
\caption{Comparison between the expected gamma-ray flux from a light scalar and the $95\%$C.L. limits  given by the {\it Fermi}-Lat collaboration, Figure 2 in \cite{Abdo:2010ex}. For the 10 GeV candidate the limits are extracted assuming annihilation into $b\bar{b}$ with a BR of 100$\%$. The limits for the 6 GeV candidate are our extrapolations, assuming BR=80\% BR in $b\bar b$ and BR=20\% in $\tau^+\tau^-$. }
\end{table}

Following a suggestion made in \cite{Fitzpatrick:2010em}, we may confront the model to data on the gamma flux from dwarf galaxies recently released by the {\em Fermi}-LAT collaboration \cite{Abdo:2010ex}.  The analysis in \cite{Abdo:2010ex} gives, for various dwarf galaxies, the 95 $\%$ C.L. limit on the total flux $\Phi$ of gamma rays (with energy between 100 MeV and 50 GeV) that may be produced through annihilation of dark matter. The published analysis, which is quite sophisticated, is limited to candidates with a mass larger than 10 GeV. However the spectrum of photons is quite similar for slightly lighter candidates (see Figure \ref{fig:spectra}), so we expect the constraints to extrapolate smoothly for, say, a 6 GeV candidate. For the sake of illustration, we consider the limits from two representative dwarf galaxies, Draco and Ursa Minor  \cite{Abdo:2010ex}. In Table I, we give the predictions for the singlet scalar model for candidates with mass 6 and 10 GeV, assuming the NFW profiles as used by the collaboration (see Table 4 of \cite{Abdo:2010ex}), and for $\sigma v = 2.5\cdot 10^{-26}$ cm$^3\cdot$s$^{-1}$. 

The 10 GeV (6 GeV) scalar interacting through the Higgs has a BR into $\tau^+\tau^-$ of order $10 \%$ ($20 \%$). For the limits on the flux, we refer to Figure 2 in \cite{Abdo:2010ex}, which display the limits obtained for a BR=100\% into $b\bar b$, and for BR=80\% into $b\bar b$ and BR=20\% into $\tau^+\tau^-$. For the 10 GeV candidate, it is a good approximation to take the limits assuming a BR of 100\% in $b\bar b$ (Table I). As shown in Figure 2 in \cite{Abdo:2010ex}, the BR$\simeq 10\%$ in $\tau^+\tau^-$ should give a limit on the flux that is ${\cal O}(10\%)$ stronger, at most. For a 6 GeV candidate, we use the figure with BR$\simeq 20\%$ in  $\tau^+\tau^-$ and BR$\simeq 80\%$ in $b \bar b$ and naively extrapolate the curve down to 6 GeV. In all cases the predicted fluxes are larger than the limits. From the numbers given in Table I,  we thus tentatively conclude that the scalar singlet model with a NFW profile may be excluded at 95 \% C.L. by data from dwarf galaxies, and  that it would be interesting to extend the analysis of \cite{Abdo:2010ex} to lighter WIMP candidates.

To close the discussion on indirect constraints, we mention the fact that the annihilation and mass of the scalar singlet candidate makes it  also a very natural candidate to solve the primordial  $^6 Li$ problem (see Figure 3 in \cite{Jedamzik:2009uy}).

\begin{figure}[t!]
\begin{minipage}[t]{0.5\textwidth}
\centering
\includegraphics[width=0.80\columnwidth]{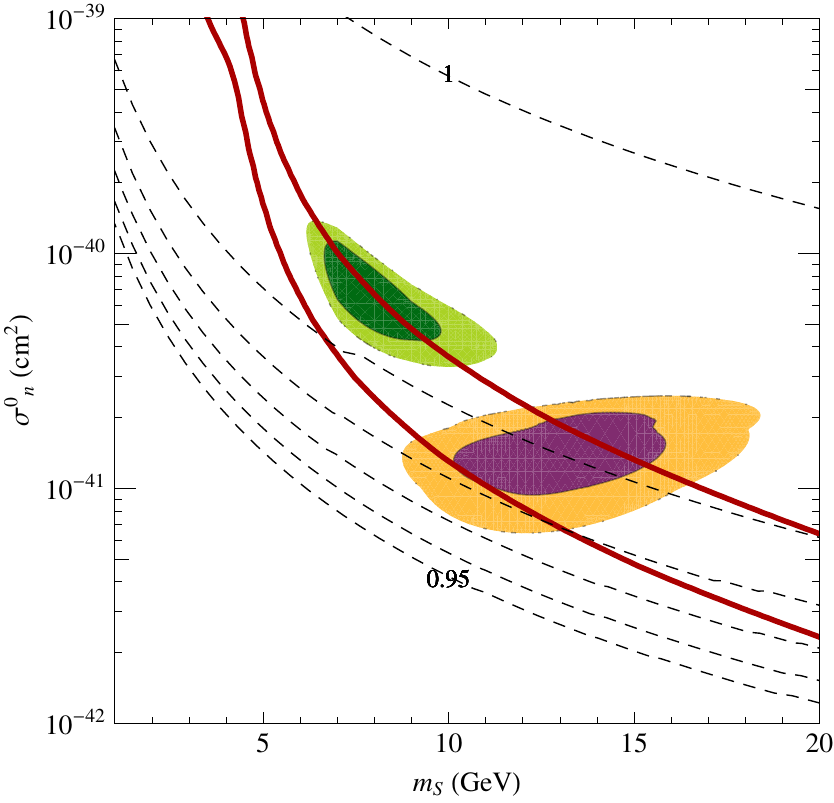}
\end{minipage}
\hspace*{-1.5cm}
\begin{minipage}[t]{0.5\textwidth}
\centering
\includegraphics[width=0.80\columnwidth]{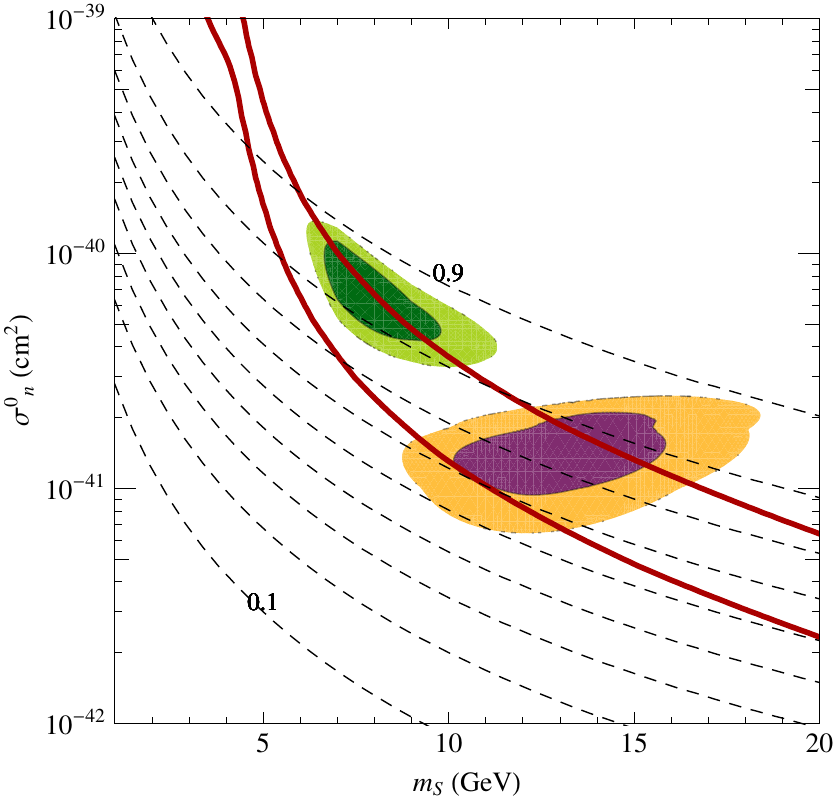}
\end{minipage}
\caption{Higgs invisible decay branching ratio for $m_h=120$ GeV (left panel) and $m_h=180$ GeV (right panel).}
\label{fig:higgsBR}
\end{figure}

Last but not least, a light WIMP in the form of a scalar coupled to the Higgs would imply that the 
Higgs mostly decays into a pair of dark matter particles~\cite{Burgess:2000yq,Barger:2007im,Andreas:2008xy,He:2009yd}. We would like to point out that the DAMA and CoGeNT regions could be distinguished from a measurement of the invisible Higgs decay branching ratio. For $m_h=180$ GeV the effect is striking: as Figure \ref{fig:higgsBR} shows, taking into account the CDMS-Si limit and the WMAP region, the DAMA region gives  $60\% \lsim BR \lsim  70\%$, while for the CoGeNT region, one has $75\%\lsim BR \lsim 90\%$. This difference is larger than the expected $\sim 10\%$ LHC  sensitivity on the invisible branching ratio \cite{Eboli:2000ze}. For $m_h=120$ GeV, the difference is much reduced, because the invisible channel largely dominates the decay width: we 
get $98\% \lsim BR \lsim 99\%$ and $BR \gsim 99\%$ for DAMA and CoGeNT respectively. 

\section{Xenon100 exclusion limits}

Shortly after CoGeNT, the Xenon100 collaboration  has released an analysis of their first data,  collected over only 11.2 days \cite{Aprile:2010um}.  The collaboration has found no event consistent with dark matter, and it has the world best SI exclusion limits for  dark matter mass  $ \lsim 80$ GeV (Figure 5 in  \cite{Aprile:2010um}).  More surprizing, at first sight, is the limit  set on lighter WIMPs, with mass $ \lsim 10$ GeV. In particular the CoGeNT and DAMA (with and without channelling) are excluded with 90\% confidence level\footnote{That something non-standard is being done in this region may be appreciated by comparing Figure 5 of  \cite{Aprile:2010um} to the preliminary results presented at recent conferences (see for instance \cite{LNGS_aprile}) and is subject to discussion \cite{Collar:2010gg,Collaboration:2010er,Collar:2010gd}.}. In this section, we present our own analysis of the low mass region, following the guidelines provided in Refs.\cite{Aprile:2010um,Collar:2010gg,Collaboration:2010er,Collar:2010gd}, to which we refer for more details. For light dark matter candidates, the relevant features of the analysis of Xenon100 are as follows. 

To discriminate dark matter (and neutron) collisions with Xe from electron and gamma events (a large source of background) the experiment relies on photoelectrons (PE) produced by scintillation (S1 signal). The relation between the mean number of PE, $n_{PE}(E_{nr})$, and the recoil energy  $E_{nr}$ (in keV)  is given by the so-called scintillation efficiency (Leff).  Which Leff to use, and how to extrapolate Leff at low $E_{nr}$ (where no measurements of Leff exist) is an important issue \cite{Collar:2010gg}. The experimental situation on Leff is summarized in Figure 1 of Ref.\cite{Collaboration:2010er}. The exclusion limit set by Xenon100 in \cite{Aprile:2010um} is based on the best fit to current experimental data (LeffMed here), which gives Leff $\approx 0.12$ at small nuclei recoil energies $E_{nr}$. Furthermore Leff is set to zero for $E_{nr}\lsim 1$ keV.  A more conservative choice (LeffMin here, corresponding to a lower 90\% C.L. fit to the data)  gives a Leff which decreases monotonically with $E_{nr}$ and vanishes at $E_{nr}\leq 1$ keV. The Zeplin experiment (also a Xe 
experiment) uses a different Leff, which is essentially zero below 6-7 keV (LeffZep here) \cite{Lebedenko:2008gb}. 

\begin{figure}[t!]
\begin{minipage}[t]{0.5\textwidth}
\centering
\includegraphics[width=0.80\columnwidth]{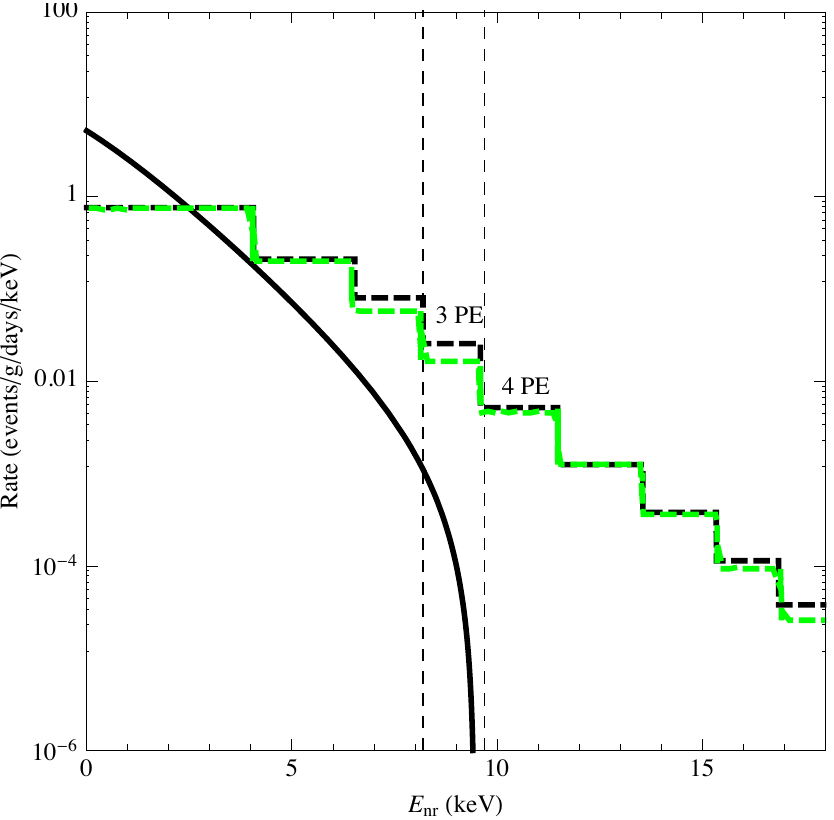}
\end{minipage}
\hspace*{-1.5cm}
\begin{minipage}[t]{0.5\textwidth}
\centering
\includegraphics[width=0.80\columnwidth]{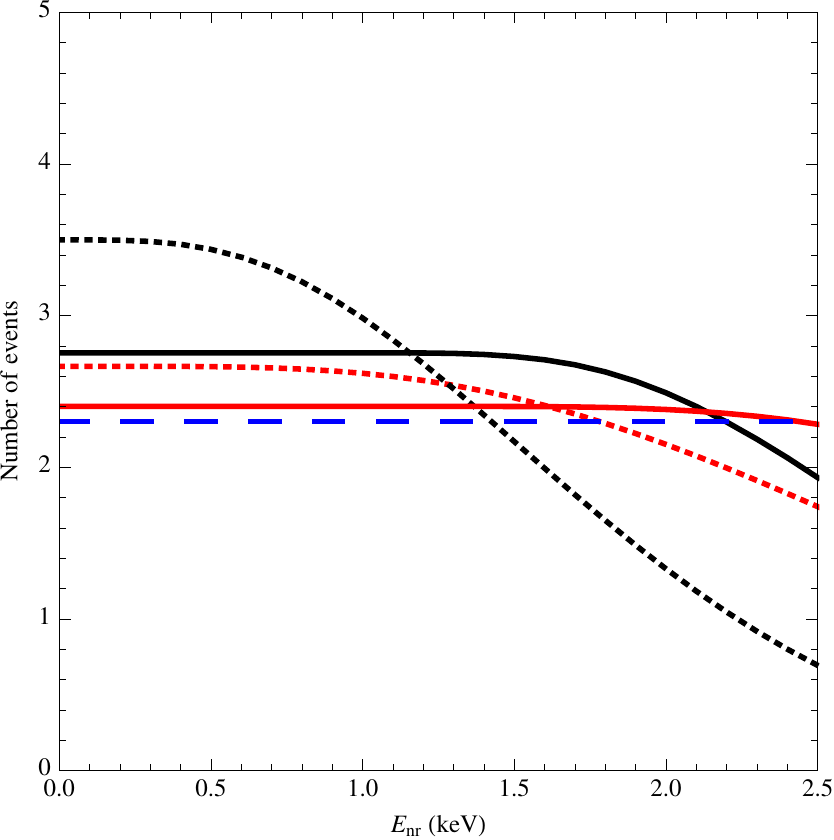}
\end{minipage}
\caption{Left panel: Effect of Poisson fluctuations on the expected signal (continuous curve) from a candidate with mass $= 10$ GeV and $\sigma_{SI} = 10^{-41}$ cm$^2$, assuming LeffMin (in green the Xenon100 bins, in black our bins -- see text).  The two vertical dashed lines correspond to the 3PE and 4PE thresholds. \\
Right panel: Effects of changing the cutoff in the recoil energy (see text) on the number of events above 3PE (in black) and 4PE (in red) for LeffMed (short dashed), LeffMin (continuous), for a candidate of $6.5$ GeV for LeffMed and of $7.5$ GeV for LeffMin, both for $\sigma_{SI} = 10^{-40}$ cm$^2$. The horizontal dashed line corresponds to the 90\% C.L. exclusion limit, which corresponds to 2.3 events according to Poisson statistics.}
\label{fig:moreXenon100}
\end{figure}

In  Figure \ref{fig:moreXenon100}, following \cite{Collaboration:2010er}, we show the theoretical event rate (per kg and per day) for a mass $= 10$ GeV and $\sigma_{SI} = 10^{-41}$ cm$^2$ candidate, typical of the DAMA region. We use $v_{esc} = 544$ km$\cdot$s$^{-1}$ as the Xenon100 collaboration. In the same Figure, the size of the bins correspond to 0PE, 1PE, etc., given here for LeffMin. The threshold depends on the acceptance of the detector. For 3PE the acceptance is about $50$\%, and about $70$\% for 4PE (Figure 3 of Ref.\cite{Aprile:2010um}). A standard analysis would give no event with 3PE or more. 
However, because of fluctuations (assumed to be poissonian) in the number of PE produced, and because the event rate is exponentially rising at low $E_{nr}$ (for elastic scattering), the rate is much larger at 3PE and 4PE, as shown by the histograms in Figure \ref{fig:moreXenon100}, obtained by  convolution of the theoretical rate with a Poisson distribution of mean $n_{PE}(E_{nr})$. For the sake of comparison we give the result of our calculation (black histogram), and that of Xenon100 (green histogram), see Figure in \cite{Collaboration:2010er}. We have slightly more events at 3PE (hence we will have slightly stronger limits), but otherwise the agreement is good. 

\begin{figure}[t!]
\begin{minipage}[t]{0.5\textwidth}
\centering
\includegraphics[width=0.80\columnwidth]{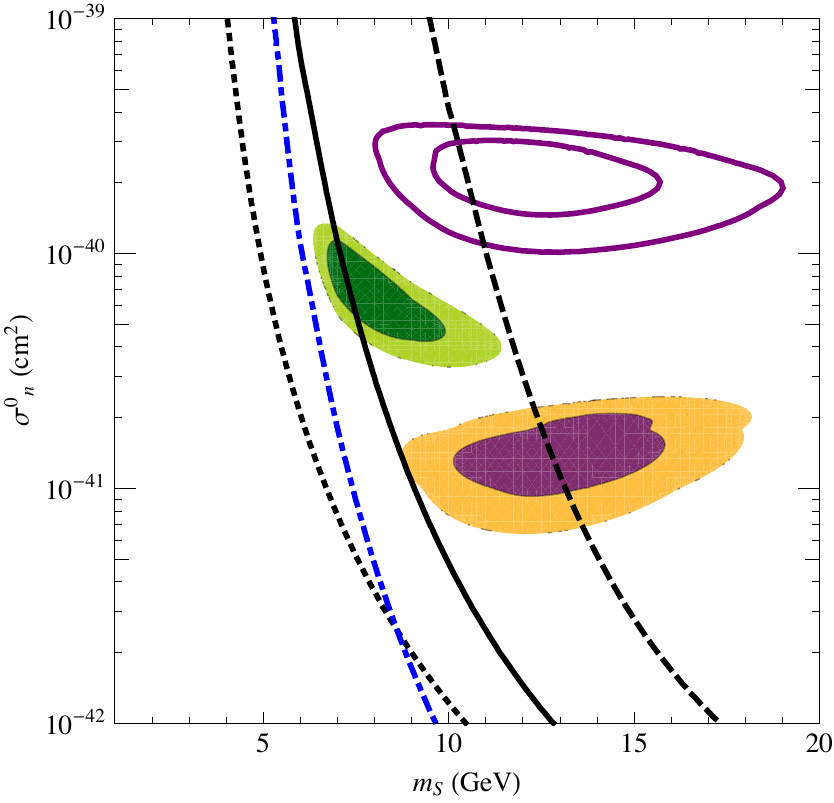}
\end{minipage}
\hspace*{-1.5cm}
\begin{minipage}[t]{0.5\textwidth}
\centering
\includegraphics[width=0.80\columnwidth]{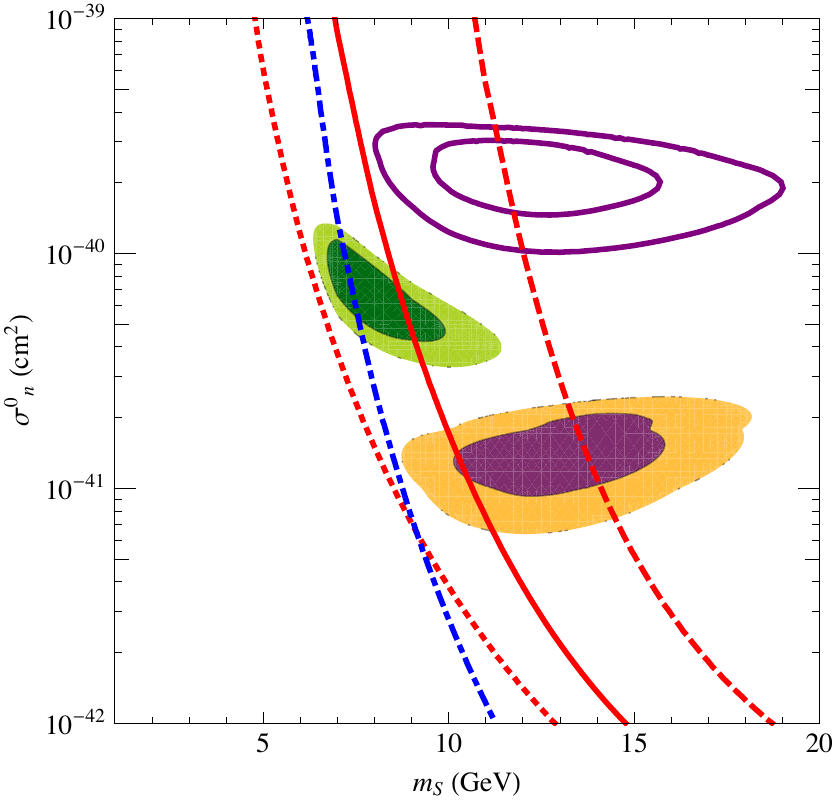}
\end{minipage}
\caption{Xenon100 exclusion limits with 90\% C.L., with threshold at 3 PE (in black, left panel) or 4 PE (in red, right panel). The curves correspond respectively to the LeffMed (short dashed), LeffMin (continuous) and LeffZep (long dashed) scintillation efficiency --- see text. For the sake of comparison,  we have taken $v_{esc} = 544$ km$\cdot$s$^{-1}$ like the Xenon100 collaboration.  The blue (dot-dashed) lines correspond to our predicted exclusion limit for Xenon100, using LeffMin and for an exposure of about 1 ton-days, assuming zero event. }
\label{fig:Xenon100limits}
\end{figure}

Our exclusion limits are shown in Figure \ref{fig:Xenon100limits}, where, in the left panel, we give the limits for a threshold at 3PE (consistent with Xenon100\cite{Collaboration:2010er}) and in the right panel the one for the more conservative choice of 4PE (consistent with Ref.\cite{Collar:2010gg}). From left to right, the exclusion limits correspond to LeffMed (short dashed), LeffMin (continuous) and LeffZep (long dashed). For the latter, the cutoff in Leff is such that the effect of Poisson fluctuations from events at low $E_{nr}$ is negligible, and one recovers the limit from a standard analysis.

An important feature we observe is that a large fraction of the events at 3PE/4PE are associated to Poisson fluctuations from the small $E_{nr}$ region,  where the rate is very large, but less than  1PE is expected on average. As a result, if one increases the $E_{nr}$ at which Leff vanishes, the number of low energy events that are lost (and which consequently do not contribute to higher energy bins) quickly increases \footnote{This loss is the reason why the number of events in the first bin is about a factor 2 smaller than the theoretical number of events.}. This effect is illustrated in the right panel of Figure \ref{fig:moreXenon100}, where we show the number of events above 3PE (resp. 4PE) for candidates with  $\sigma_{SI} = 10^{-40}$ cm$^2$, both for LeffMed and LeffMin, but changing the cutoff $E_{nr}$  at which they are assumed to vanish. The effect is most sensitive for LeffMed, where for a cutoff at $E_{nr}=1$ keV the candidate is excluded at 90\% C.L., while it is allowed if the cutoff is at $E_{nr}= 1.5$ keV. In our exclusion limits we use the function LeffMed and LeffMin with a cutoff at $E_{nr}= 1$ keV. 

In our opinion, the method followed by the Xenon100 collaboration is sound, but is very sensitive (as discussed in \cite{Collar:2010gg,Collaboration:2010er,Collar:2010gd}) to the choice of Leff at low $E_{nr}$. Here we would like to emphasize that not only the shape of Leff, but also the $E_{nr}$ at which Leff is cutoff, is a critical issue in setting exclusion limits. This being said, one must admit that  there is a tension between the CoGeNT (and {\em a fortiori} DAMA) regions and the limits set by the Xenon100 experiment, and this with only limited exposure. Assuming that {no} events are seen in the data, we give in Figure \ref{fig:Xenon100limits} our predicted exclusion limit of Xenon100, using LeffMin, for an exposure corresponding to 1 ton-days, as reported in \cite{LNGS_aprile} .

\section*{Acknowledgments}

We thank Thomas Schwetz, Karsten Jedamzik, Juan Collar and Martin Casier for useful discussions. 
Our work is supported by the FNRS-FRS, 
the IISN and the Belgian Science Policy (IAP VI-11).


\bibliographystyle{h-physrev3}
\bibliography{biblio}

\end{document}